\documentclass[a4paper]{article}

\usepackage{itgspeech2023}    
\usepackage{times}            
\usepackage[english]{babel}   
\usepackage[ansinew]{inputenc}
\usepackage[T1]{fontenc}      
\usepackage[sort&compress,numbers]{natbib}	
\usepackage{amsmath,amssymb}
\usepackage{graphicx}
\usepackage[colorlinks=false,pdfborder={0 0 0}]{hyperref}
\usepackage{units}

\title{Compression of end-to-end non-autoregressive image-to-speech system for low-resourced devices}

\author{Gokul Srinivasagan $^{1,2}$, Michael Deisher$^2$, Munir Georges$^{3,4}$}

\address{
  $^1$Saarland University, Saarbrucken, Germany\\
  $^2$Intel Corporation, Hillsboro, Oregon, USA\\
  $^3$Intel Labs, Munich, Germany\\
  $^4$Technische Hochschule Ingolstadt\\
  \href{mailto:gokulsrinivasagan@gmail.com,michael.deisher@intel.com,munir.georges@intel.com}{gokulsrinivasagan@gmail.com,\{michael.deisher,munir.georges\}@intel.com}}

\begin{document}

\maketitle

\begin{abstract}
People with visual impairments have difficulty accessing touchscreen-enabled personal computing devices like mobile phones and laptops. The image-to-speech (ITS) systems can assist them in mitigating this problem, but their huge model size makes it extremely hard to be deployed on low-resourced embedded devices. In this paper, we aim to overcome this challenge by developing an efficient end-to-end neural architecture for generating audio from tiny segments of display content on low-resource devices. We introduced a vision transformers-based image encoder and utilized knowledge distillation to compress the model from 6.1 million to 2.46 million parameters. Human and automatic evaluation results show that our approach leads to a very minimal drop in performance and can speed up the inference time by 22\%.

\hfill

\end{abstract}

\sloppy

\section{Introduction}
\label{sec:intro} 

In recent years, several advancements have been made in touchscreen-enabled personal computing devices such as cell phones, tablets, and laptops. Despite these advancements, people with visual impairments continue to have difficulty accessing these devices. Screen readers can help users to some extent. But they are OS-dependent and do not function on the blue screen of death (BSOD) or pre-boot environments. They also have limitations on legacy applications and dealing with text embedded in images.

One solution to overcome this problem is having an optical character recognition (OCR) or image-to-text module for extracting text and a text-to-speech (TTS) module for generating audio from the text. But having the two independent modules in the pipeline leads to the propagation of errors from one module to another. Moreover, optimizing the modules individually makes the system sub-optimal. Also, auto-regressive text-to-speech models incur a delay that adds to the total inference time. A non-autoregressive end-to-end image-to-speech (ITS) addresses both inference time and accuracy challenges. Earlier work \cite{chen2023end} introduced the first end-to-end non-auto-regressive neural ITS model. It was designed to be deployed on an independent embedded subsystem, thereby making it robust to OS failures. But the model was not compact and there is an opportunity to significantly reduce power and die size (cost) by further reducing the size of the model.



In order to achieve this, in this work, we try to reduce the memory footprint of the models without a major impact on the accuracy. We employ knowledge distillation to reduce the size of the ITS model. We also incorporate the latest vision transformer-based scene text recognition model for improving the accuracy of the ITS system. Having a smaller model also leads to reduced inference time, thereby decreasing the delay between the user touching a word on the screen and generation of the corresponding audio by the system.

\section{Background and Related Work}
\label{sec:related_works}
Optical character recognition (OCR) and text-to-speech (TTS)
are two domains of research, which have seen ground-breaking developments in recent years. There is a rise in the development of small non-autoregressive TTS models \cite{ren2019fastspeech, ren2020fastspeech, ren2021portaspeech}, which can generate high-quality audio in a few milliseconds. But, the image-to-speech domain is still under-explored. We follow the approach suggested by Chen et al. \cite{chen2023end} for developing an efficient non-autoregressive end-to-end ITS system.

Transformer-based models have produced state-of-the-art results not only in natural language processing but also in computer vision tasks. The vision transformer \cite{dosovitskiy2020image} has proven to perform better than advanced CNN-based models with significantly lower computational resource requirements during training. This motivated researchers to adapt transformer-based models to other computer vision tasks \cite{carion2020end}. Since vision transformer-based scene text recognition (ViTSTR) models have shown better performance than other similar-sized CNN models on OCR tasks  \cite{atienza2021vision}, we utilize these models as an image encoder. 


Several model compression approaches are used specifically for the compression of large-scale transformer-based models. Some of the popular approaches include i) knowledge distillation, ii) quantization, and iii) pruning. Knowledge distillation \cite{hinton2015distilling} involves the transfer of knowledge from a large teacher model to a small student model to mimic the functioning of the teacher. The knowledge distillation can be performed with the model's intermediate outputs, final outputs, or both. Knowledge distillation from output logits forces the student to replicate the teacher models' output, whereas intermediate knowledge distillation forces the student model to replicate the individual layer of the teacher. Quantization \cite{zafrir2019q8bert} involves representing the model weights with a reduced number of bits. It can also reduce the model size and inference time if the underlying computation device is optimized for low-precision operations. Pruning \cite{frankle2018lottery} is another popular model compression approach, which focuses on identifying and removing the less important components or weights from the model to improve performance. We wanted our ITS system to be as small as possible with comparative performance, which can be aided by knowledge distillation. Several works have proven the effectiveness of the knowledge distillation approach by achieving greater compression ratios than other techniques (\cite{sanh2019distilbert}, \cite{sun2020mobilebert} \& \cite{jiao2019tinybert}). So, in our work, we used offline knowledge distillation where the weights of the teacher are frozen, and only the student weights are updated during training.



\section{Methodology}
\label{sec:method}

\begin{figure}
\centering
\includegraphics [scale=0.6]{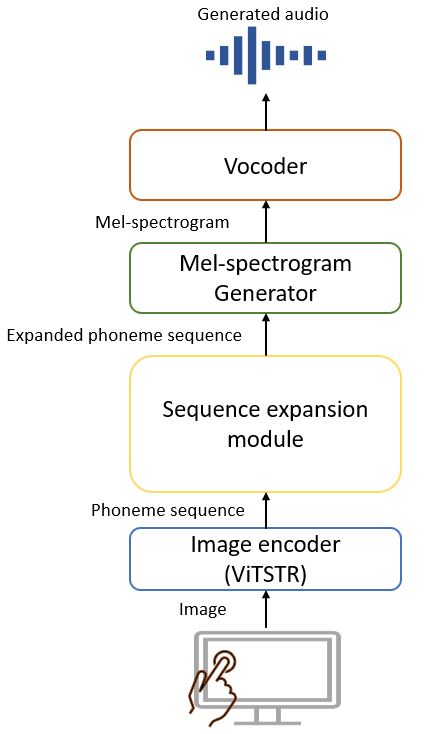}
\caption{Individual components of the image-to-speech system. The selected region from the touchscreen device is the input and the synthesized audio signal is the output.}
\label{fig:its_arch}
\end{figure}

\subsection{Image-to-speech system}

Our end-to-end image-to-speech system consists of four main components. Figure \ref{fig:its_arch} shows all the building blocks of our ITS system. The selected region from touchscreen devices is cropped and that image is given as input to our ITS system, which produces mel-spectrograms as output. The mel-spectrograms are used by the vocoder to generate audio signals. Individual blocks of our ITS system are discussed below:

\begin{description}
  \item[$\bullet$ Image encoder:] The region of the display near the user's finger or pointer is taken as input by the image encoder. Unlike the standard OCR model, where the model identifies the textual characters embedded in the image, our model is modified to predict the phoneme sequence. This is because the grapheme-to-phoneme module in conventional TTS systems is not back propagatable \cite{chen2023end}. A special token $\epsilon$ is added at the end to expand the dimension of phoneme sequence to a fixed length. The image encoder in our baseline \cite{chen2023end} uses a CNN-based model HBONet \cite{li2019hbonet} with a rectifier for normalizing the image. This block has around 3.87 million parameters accounting for 63.4\% of the model parameters. We removed the image rectifier and introduced a data augmentation step during training which generates random cropped images with different orientations. This data augmentation step helps the model to deal with un-normalized images during inference. We also incorporated a vision transformer-based OCR model \cite{atienza2021vision} into the system to decrease the phoneme error rate.

  
  \item[$\bullet$ Sequence expansion module:] The sequence expansion module takes the sequence of phonemes as input, identifies the duration (n) of each phoneme, and expands each hidden representation n times.
  
  \item[$\bullet$ Mel-spectrogram generator:] The mel-spectrogram generator takes the expanded hidden representation from the sequence expansion module and generates mel-spectrograms as output. We followed the variational autoencoder (VAE) based model \cite{ren2021portaspeech} to produce the mel-spectrogram.
  
  \item[$\bullet$ Vocoder:] The vocoder is trained separately and it takes the output from the mel-spectrograms generator to produce audio signals. The HiFiGAN v2 \cite{kong2020hifi} is used as a vocoder in the system due to its compact size of 0.9 million parameters.
  
\end{description}

\subsection{Image encoder}

\begin{figure*}[ht]
\centering
\includegraphics[width=\textwidth]{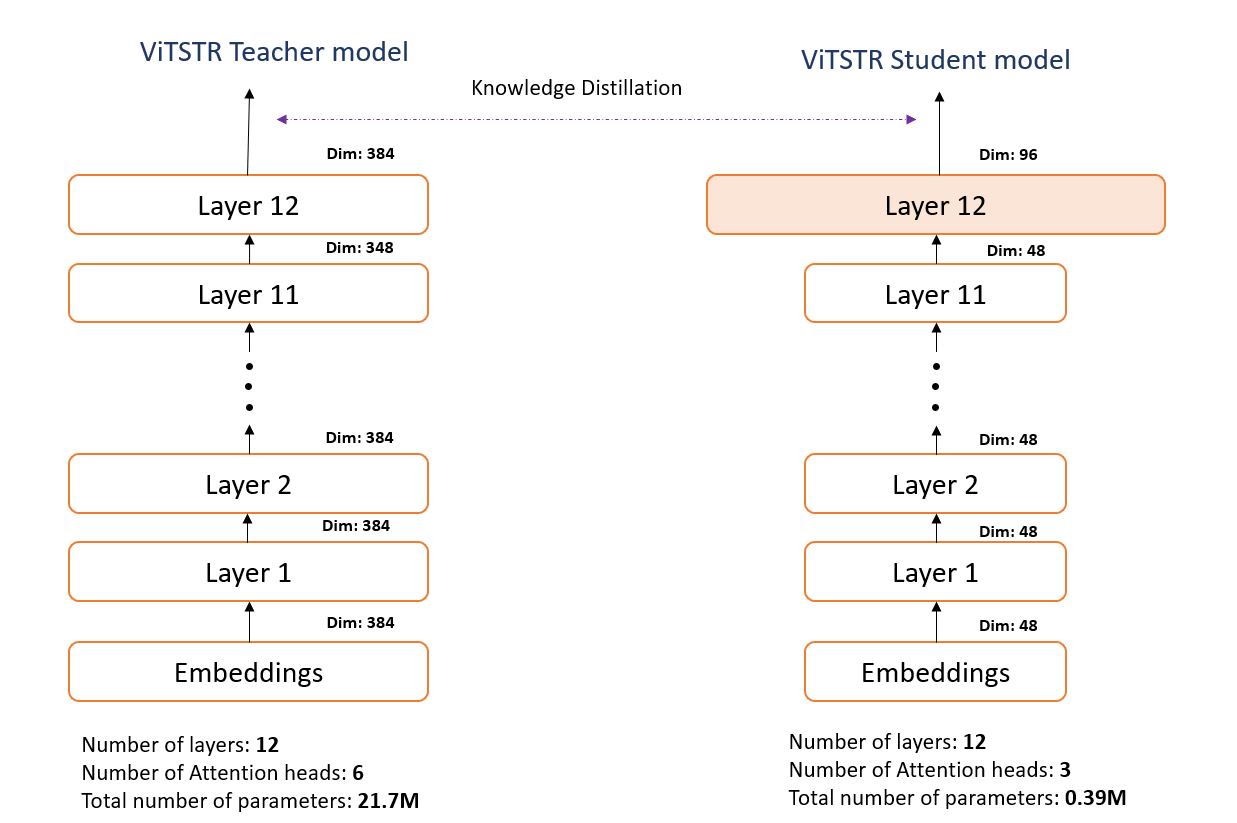}
\caption{Architecture of the teacher and student image encoder models. The student model has a hidden dimension of 96 in the last layer and 48 in the rest of the layers.}
\label{fig:experiment}
\end{figure*}

Figure \ref{fig:experiment} shows the overall architecture of the ViTSTR-based teacher and student models. Both teacher and student models have 12 layers but they differ in the size of the hidden dimension. 


\subsubsection{Teacher model}
We used the ViTSTR model with a hidden dimension of 384 as the teacher model. The model is initialized with weights from DeiT \cite{touvron2021training} and trained on the downstream task for identifying phonemes from the image. The model is trained using the data from the MJSynth (MJ) \cite{jaderberg2014synthetic} and SynthText (ST) \cite{gupta2016synthetic} datasets.

\subsubsection{Student model}
For the student model, we focused on reducing the width because it provides greater compression than reducing the depth of the network. Our experiments also showed that the reduction in the hidden dimension of the model gives better performance than compressing the number of layers, revealing that a larger number of layers is crucial for the higher accuracy of the model.

Transformer-based models usually have repeating blocks of n-layers stacked on top of one another. For our student model, we double the hidden dimension of the final layer only to 96 while keeping the rest of the layer dimension to 48. This modification gave a large gain in accuracy with a minimal increase in the number of parameters. We also did several experiments varying the hidden dimensions and number of layers. Using hidden dimensions smaller than 48 or using less than 12 layers greatly degrades accuracy. Also, we experimented with increasing the last three layers of the image encoder to 96 but it did not improve accuracy. So, we decided to only increase the dimension of the final layer.

We proposed two variations of the model. In the compressed end-to-end ITS model v1, we focused on making the image encoder efficient by introducing a vision transformer and removing the rectifier. The compressed end-to-end ITS model v2 has 50\% fewer decoder layers at the mel-spectrogram generator than v1.


\section{Experimental results}
\label{sec:results} 

\begin{table*}[!ht]
    \centering
    \begin{tabular}{|l|l|l|}
    \hline
        \textbf{Model} & \textbf{Inference time (ms/image)} & \textbf{Number of parameters} \\ \hline
        \textbf{End-to-end ITS model (baseline)} & 45 ms &  6.1 M \\ \hline
        \textbf{Compressed end-to-end ITS model v1} & 38 ms & 2.63 M \\ \hline
        \textbf{Compressed end-to-end ITS model v2} & \textbf{35 ms} & \textbf{2.46 M} \\ \hline
    \end{tabular}
    \caption{Comparison of size and inference time of various image-to-speech models.}
    \label{tab:size_reduction}
\end{table*}

\begin{table*}[!ht]
    \centering
    \begin{tabular}{|l|l|l|}
    \hline
        \textbf{Model} & \textbf{Phoneme Error rate (PER)} & \textbf{Word accuracy} \\ \hline
        \textbf{End-to-end ITS model (baseline)} & 8.82\% & 81.26\% \\ \hline
        \textbf{Compressed end-to-end ITS model v1} & 9.46\% & 79.16\% \\ \hline
        \textbf{Compressed end-to-end ITS model v2} & 10.57\% & 78.5\% \\ \hline
    \end{tabular}
    \caption{Comparison of performance of various image-to-speech models}
    \label{tab:performance}
\end{table*}

\subsection{Implementation details}
For the image encoder, we resize all the images in the dataset to 224x224. The maximum length of the output phoneme sequence is set to 25. In each batch of training, 50\% of the samples come from the MJ dataset and the remaining 50\% are from the ST dataset. A grapheme-to-phoneme package \cite{g2pE2019} is used for converting text labels to phoneme sequences in the training dataset. The models were trained for 300,000 steps with a batch size of 96. Cross-entropy is used for training the teacher model. We employ data augmentation techniques to train both teacher and student models. Gradient clipping with value 5 is also used. The Adam optimizer with a learning rate of 5e-5 is used for training the teacher model. For the student, a learning rate of 5e-4 is used. The image encoder is trained to identify 72 phoneme classes. The model is trained to identify text from the English language only.

For training the student, cross-entropy is used as the loss function between the student output and ground truth labels. Mean squared error (MSE) loss is used between teacher and student outputs. For knowledge distillation, we give equal importance to both these losses by selecting an alpha value of 0.5. We also use gradient cosine similarity \cite{georges2020compact} if the gradients have non-negative cosine similarity with the target gradient. Otherwise, the knowledge distillation loss is not considered and the model is trained with only cross-entropy loss.  We performed offline knowledge distillation with only the final outputs in our experiments. Due to the mismatch in intermediate hidden dimensions of teacher and student models, we did not perform intermediate layer knowledge distillation in our experiments.

The dataset for our ITS system is generated with ground truth audio from the Azure TTS service by taking 50\% of images from the MJ dataset. The resulting dataset contains 658,583 images and corresponding audio. L1 loss and structural similarity loss are used to train the mel-spectrogram generator. Mean square error (MSE) is used for training the duration predictor. All our experiments are carried out using Nvidia RTX 2080 ti GPU.

\subsection{Training procedure}
Training is performed in four different stages. First, the teacher image encoder model is trained. Then the weights of the teacher model are frozen and the student model is trained using knowledge distillation from the teacher model. Subsequently, the student model is used as an image encoder by freezing its weights, and the entire ITS system is trained. Finally, image encoder model weights are unfrozen, and the entire ITS system is trained for a few steps.

\section{Evaluation}
For evaluating the performance of the compressed end-to-end ITS system, we used both human and automatic evaluation metrics. The evaluation dataset is synthetically generated by selecting the 3000 most frequently occurring words from Wikipedia. These images were generated to have different fonts, font colors, sizes, and background colors. 

For human evaluation, we selected a small group of 15 participants. The participants were asked to compare the output of compressed and uncompressed end-to-end ITS systems using 10 randomly selected audio outputs. The results of the listener test show that outputs from both systems were rated equally good for 61.3\% of cases. Whereas in 20\% of cases, the participants preferred the uncompressed ITS output, and in the rest of the cases the compressed ITS system was preferred.

For the automatic evaluation of our ITS system, we use an automatic speech recognition (ASR) model to convert the generated audio to text and evaluate it with ground-truth labels. We use a Wav2vec2 large model fine-tuned on Librispeech \cite{7178964} and TED talks (TED-LIUM) \cite{hernandez2018ted} datasets for transcribing the audio. The phoneme error rate and word accuracy are used as performance evaluation metrics. 

\section{Discussion}
Table \ref{tab:size_reduction} highlights the reduction in the size of our ITS system and the improvement in inference time by our approach. Our compressed ITS v2 model has 67.2\% fewer parameters than the non-end-to-end ITS model \cite{chen2023end} and produces a result 46\% faster. When compared with the baseline ITS model \cite{chen2023end}, our model is 59.7\% more parameter efficient and 22.2\% faster. 

Table \ref{tab:performance} shows the comparison in accuracy and phoneme error rate of our compressed ITS models. The results show that both our models have a 2-3\% drop in accuracy and a 1-2\% increase in phoneme error rate. Using the proposed approach, we attained a compression of more than 60\% with a very minimal drop in accuracy. Also, removal of the rectifier had a minimal impact on the model performance for the data sets used here.  This is likely due in part to the improved data augmentation compared to the previous work.

Some suggestions to further improve the performance of the ITS system include improving the quality of the generated audio by vocoder \cite{du2022vqtts, 10095298, 9632770} and jointly training the vocoder along with our ITS system \cite{lim2022jets}. Improving the vocoder efficiency is another research direction \cite{9746675, srivastava2023fast}.

\section{Conclusion}
\label{sec:conclusion}
The proposed image encoder model and knowledge distillation resulted in the compression of the ITS model by 59.7\%, reducing the number of parameters to 2.46M. This smaller-sized ITS model is more suitable for low-resource devices and reduces the inference time by 22.22\%. Currently, the proposed ITS system is limited to English alphabetic characters. In the future, we plan to extend the functionality of the ITS system to support non-English character sets, abbreviations, acronyms, word overlaps and incomplete word selections. 




\small
\bibliographystyle{ieeetr}
\bibliography{example}


\end{document}